\documentclass[fleqn,usenatbib]{mnras}

\usepackage{newtxtext,newtxmath}

\usepackage{caption}  % To make xtabular captions match
\usepackage[T1]{fontenc}

\DeclareRobustCommand{\VAN}[3]{#2}
\let\VANthebibliography\thebibliography
\def\thebibliography{\DeclareRobustCommand{\VAN}[3]{##3}\VANthebibliography}

\usepackage{graphicx}	% Including figure files
\usepackage{amsmath}	% Advanced maths commands

\usepackage{hyperref} % make references, figures, and tables clickable!
\hypersetup{
  colorlinks = true, %Colours links instead of ugly boxes
  urlcolor = blue, %Colour for external hyperlinks
  linkcolor = blue, %Colour of internal links
  citecolor = blue %Colour of citations
}

\newcommand{\textcode}[1]{\texttt{#1}} % To make words look like code

\title[\textit{Gaia}, NED, and SIMBAD in nearby galaxies]
{Comparing \textit{Gaia}, NED and SIMBAD source classifications in nearby galaxies} 

\author[Hales \& Barmby]{ 
J. Hales$^{1}$\thanks{(E-mail: jhales5@uwo.ca)} and
P. Barmby$^{1,2}$\thanks{(E-mail: pbarmby@uwo.ca)}
\\
$^{1}$Department of Physics \& Astronomy, Western University, London, Canada, N6A 3K7\\
$^{2}$Institute for Earth and Space Exploration, Western University, London, Canada, N6A 3K7
}

\date{Accepted 2024 August 22. Received 2024 August 13; in original form 2023 March 24
}

\pubyear{2024}

\begin{document}
\label{firstpage}
\pagerange{\pageref{firstpage}--\pageref{lastpage}}
\maketitle

\begin{abstract}
\textit{Gaia} Data Release 3 (DR3) provides the first classifications for the sources in \textit{Gaia}'s all-sky database. Most \textit{Gaia} sources are stars in the Milky Way, but DR3 also contains many sources that belong to nearby galaxies, as well as background galaxies and quasars. In this work, we compare the \textit{Gaia} classifications from the Discrete Source Classifier (CU8-DSC) module to the more detailed and heterogeneous classifications in NED and/or SIMBAD for sources with sky positions within twice the Holmberg radius of nearby galaxies.
Matching these catalogues gives approximately $3.2\times10^5$ unique \textit{Gaia} matches for $4\times10^5$  sources over 1040 galaxies (excluding some large Local Group galaxies) in the Local Volume Galaxy catalogue. Matched sources contain a lower fraction of \textit{Gaia}-classified stars and higher fractions of galaxies and quasars ($\sim95$, 2 and 2~per~cent, respectively) than DR3 overall. Considering NED (SIMBAD) classifications as truth values, the balanced accuracy of \textit{Gaia} classification is 0.80 (0.83): the most common disagreements are literature-classified galaxies  \textit{Gaia}-classified as stars and literature-classified stars \textit{Gaia}-classified as quasars. 
Purity ($P$) and completeness ($C$) metrics show that
agreement between \textit{Gaia} classification and NED/SIMBAD classification is best for stars ($P,C\sim 0.9$), and decreases for quasars ($P<0.3$, $0.7<C<0.8$), galaxies ($0.7<P<0.8$, $0.3<C<0.6$), white dwarfs
($0.04<P<0.6$, $C\sim 0.6$), and binary stars ($P,C<0.1$). NED or SIMBAD sources classified only by detection wavelength are most often \textit{Gaia}-classified  as stars, while non-stellar components of galaxies appear in all  \textit{Gaia} classes. 

\end{abstract}


\section{Introduction}

Classifying astronomical sources is a complex 
process: classifications can change as new observations, new physical understanding, and new statistical methods
become available.
Extragalactic astronomy relies on various databases to combine and organize measurements. Two of the largest of these databases are the NASA Extragalactic Database (NED) and the Set of Identifications, Measurements and Bibliography for Astronomical Data (SIMBAD), both of which are compilations of results from the literature.
SIMBAD was designed for individual star properties before being updated to include galaxies \citep{ref:simbad_original, ref:simbad_update}. 
It contains 13.41 million distinct sources and 53.73 million identifiers
categorised using a detailed hierarchical classification scheme.
SIMBAD also contains `candidate' classifications (e.g. `galaxy candidate'). 
NED was designed to catalogue sources outside of the Milky Way \citep{ref:ned_original} and contains over 
1.11 billion distinct sources, 13.96 billion photometric data points, and 502 million classifications 
as of data release 32.3.1\footnote{\label{footnote:ned_CH}\href{https://ned.ipac.caltech.edu/CurrentHoldings}{https://ned.ipac.caltech.edu/CurrentHoldings}}.  NED uses a flat classification scheme, with no hierarchy, that assigns sources to more general types tailored to the extragalactic regime.

The heterogeneity of literature-based databases and their classification schemes result in discrepancies in source classifications. 
\citet{ref:comparing_NS} compared the classifications between NED and SIMBAD for sources in the nearby galaxy regime, finding
that $\sim66$~per~cent of classifications showed strong agreement, 22~per~cent showed weak agreement (e.g. `ultraviolet source' to `star') and 12~per~cent did not agree.
Many of the weakly-agreeing classifications were for sources whose exact nature is unknown, for example those classified only by detection wavelength.
\citet{ref:comparing_NS} noted that ``many sources\dots are contained in only one of NED or SIMBAD'' and recommended the use of both databases for the most complete picture of an individual galaxy's components.

The European Space Agency’s \textit{Gaia} satellite \citep{ref:Gaia_mission_2016} is in the process of creating one of the largest and most complete all-sky surveys. The third and most recent data release, the \textit{Gaia} Data Release 3 (DR3), was released in two parts. 
The first installment, the Early Data Release 3 (EDR3) provides homogeneous astrometry and photometry for 1.81 billion sources brighter than $G\approx21$ mag collected during the first 34 months of the satellite's operation \citep{ref:EDR3, ref:EDR3_photometry}. 
The EDR3 contains complete astrometry (position, parallax, and proper motion) for 1.47 billion sources, with five-parameter astrometry for 585 million sources and six-parameter astrometry for 883 million sources, encompassing 93~per~cent of sources brighter than $G\approx17$ mag \citep{ref:EDR3_astrometry}. 
Mean positions are available for the remaining 343 million sources \citep{ref:EDR3_astrometry}.
The second installment provides supplementary data, including source classification probabilities computed by the Data Processing and Analysis Consortium for 1.59 billion sources  \citep{ref:DR3_source_classification}. 

The completeness and uniformity of the \textit{Gaia} data provide a new comparison standard for source classifications in literature-based databases.
Several recent works have crossmatched \textit{Gaia} data with the literature for specific source types. \citet{ref:lit_validating_YSOs} crossmatched the sample of young stellar object candidates identified by \textit{Gaia} with catalogues of young stellar objects in the literature to validate \textit{Gaia} classifications. 
\citet{ref:lit_DR3_variable_objects} compiled \textit{Gaia} DR3 data for various types of known variable objects in the literature. \citet{ref:lit_DR3_hotsubdwarfs} used \textit{Gaia} EDR3 data to obtain astrometry and photometry for spectroscopically classified hot subluminous stars in the literature and identify new candidates.
\citet{ref:lit_DR3_variable_AGN} identified new variable active galactic nuclei candidates using only \textit{Gaia} DR3 data. \citet{ref:lit_DR2_globular_clusters} used \textit{Gaia} DR2 data to create a catalogue of globular cluster candidates in NGC~5128.

In this paper, we compare \textit{Gaia} classifications for sources in the vicinity of nearby galaxies with those from NED and SIMBAD. 
We follow previous work that 
estimated the foreground content of  \textit{Gaia} DR3 in the vicinity of nearby galaxies \citep{ref:gaia_foregrounds} in
using the Local Volume Galaxy Catalogue \citep{ref:LVG} as a parent sample.
We create an inventory of sources in nearby galaxies from the literature crossmatched with \textit{Gaia}, including classifications from the DR3. 
We analyse the degree to which \textit{Gaia} classifications agree with the literature and further classify sources given ambiguous and candidate classifications, with the goal of understanding how \textit{Gaia} classifications might be used in the study of nearby galaxies.

\section{Methods}
\label{sec:methods}

\subsection{Galaxy sample and source retrieval}
\label{sec:glx_sample}

We used the June 2022 version of the Local Volume Galaxy catalogue \citep[][hereafter LVG]{ref:LVG} for our sample of nearby galaxies. The LVG is an all-sky catalogue containing entries for 1421 galaxies within 11~Mpc of the Milky Way or with radial velocities $<600$~km~s$^{-1}$, excluding the Virgo cluster.
A few Local Group galaxies in the LVG were omitted from our sample, because of their large angular extent or because they have already been the target of individual studies with \textit{Gaia}: the Milky Way, M31, M32, M33, the Large and Small Magellanic Clouds, and the Sagittarius dSph.
As a measure of each galaxy's size, we used the Holmberg radius, the length of the semi-major axis corresponding to the $B=26.5$ isophote. 
We adopted the coordinates and angular diameter corresponding to the Holmberg isophote as listed in the LVG for all galaxies to ensure consistency.
Galaxies without visible light size measurements were omitted from the sample.

We retrieved all sources within the Holmberg radius of the centre of each galaxy from NED and SIMBAD using the open-source command line tool \textcode{galaxy\_data\_mines}\footnote{\label{footnote:gdmines}\href{https://github.com/mshubat/galaxy\_data\_mines}{https://github.com/mshubat/galaxy\_data\_mines}}. 
\textcode{galaxy\_data\_mines} provides an efficient way to crossmatch common sources between NED and SIMBAD and determine the degree to which classifications agree. 
For a comprehensive description of the algorithm used, we refer the reader to \citet{ref:comparing_NS}.
Following that work, we used a tolerance of 5~arcseconds for crossmatching sources between NED and SIMBAD. While this is larger than the \textit{Gaia} astrometric precision, the astrometric precision of literature data from NED and SIMBAD  is likely worse and certainly more heterogeneous. Tests by \citet{ref:comparing_NS} indicated that this tolerance captured most matches without significant spurious matching.
Galaxy names are often inconsistent between NED, SIMBAD, and the LVG: to mitigate this, we searched each database using the sky coordinates of the centre of each galaxy as listed in the LVG. 
We retained sources that appear in only one of NED or SIMBAD to obtain a complete catalogue for all sources in the literature in each galaxy.
1285 galaxies in the LVG have one or more NED or SIMBAD sources within the Holmberg radius.
Median values for these galaxies are: distance 7.7~Mpc, radius 1.8~kpc, and absolute $K$-magnitude $-20.7$.
Most are dwarf irregulars (T-types 8--10) or lenticulars (T-types $-1$ to $-3$).

For a description of the astrometric and photometric processing of the \textit{Gaia} EDR3, we refer the reader to \citet{ref:EDR3_astrometry} and \citet{ref:EDR3_photometry}.
We used the dataset from \citet{ref:gaia_foregrounds}, which provides all \textit{Gaia} DR3 astrometry and photometry available for sources within double the Holmberg radius from the centre of each galaxy and identifies foreground sources based on astrometric properties: 
1092 galaxies in the LVG have one or more \textit{Gaia} DR3 sources within the Holmberg radius.
Median values for these galaxies are: distance 7.2~Mpc, radius 2.2~kpc, and absolute $K$-magnitude $-21.1$,
with a similar distribution of morphological types to the galaxies with NED and SIMBAD sources. 
Table~\ref{table:src_counts} summarises the filtering process.

\begin{table}
\centering
\caption{Galaxy and source sample size after down-selection} 
\label{table:src_counts}
\begin{tabular}{lrr} 
\hline
& $N_{\rm glx}$ & $ N_{\rm src}$\\
\hline
\hline
All Local Volume Galaxies (parent sample)  &  1421 & \\
with tabulated sizes & 1411 &\\
 with large galaxies removed & 1401 & \\
 with NED and/or SIMBAD sources  &  1285 &  1360173\\
with \textit{Gaia} sources  &  1092 & 493739\\
with \textit{Gaia} and (NED and/or SIMBAD) sources &  1040 & 404173\\
\hline
\end{tabular}
\end{table}

\subsection{Classifying \textit{Gaia} sources}
\label{sec:gaia_classifications}

The \textit{Gaia} Discrete Source Classifier (CU8-DSC) module uses machine learning to compute the probabilities of each \textit{Gaia} source to be a star, galaxy, quasar, physical binary star, and white dwarf. 
We refer the reader to \citet{ref:DR3_source_classification}, \citet{ref:DR3_extragalactic_content}, and \citet{ulla2022} for a complete description of the algorithms and training data. 
Briefly, the classification methodology in DSC involves three algorithms: 
Allosmod  \citep{bailerjones2019}, which uses astrometric and photometric measurements to classify using a Gaussian mixture model;
Specmod, which uses the BP/RP spectra with an ensemble of classification trees; and
Combmod, which combines the probabilities from Allosmod and Specmod using a Bayesian approach.
Class probabilities are provided in the \textcode{astrophysical\_parameters} table in the \textit{Gaia} archive.
We used \textcode{classprob\_dsc\_combmod}, and following the recommendations by \citet{ref:DR3_source_classification}, we assigned each source the classification with the highest combined probability above 50~per~cent.
If no probability exceeded 50~per~cent the source was listed as unclassified.
The sample classified as stars by \textcode{classprob\_dsc\_combmod} has $>99$~per~cent completeness and purity,%
\footnote{In the classification and machine-learning literature, completeness and purity are more commonly called recall and precision, respectively.}
although the dominance of stars in the sample means that these metrics are not particularly meaningful \citep{ref:DR3_source_classification}. 
While \textcode{classprob\_dsc\_combmod}-classified galaxy and quasar samples have high completeness (0.94 and 0.92 respectively), they have rather low purity (0.22 and 0.24); purity improves to about 0.4 at higher Galactic latitudes. 
As described by \citet{ulla2022}, these metrics are adjusted to account for class weights and priors, so are affected by the rarity of non-stars in the full dataset.

\subsection{Crossmatching NED and SIMBAD sources to \textit{Gaia} sources}
\label{sec:gaia_sources}

The combined list of NED and/or SIMBAD sources in the vicinity of LVG galaxies contains approximately 1.4 million sources
(Table~\ref{table:src_counts}), selected solely on sky position.
A NED/SIMBAD crossmatch was performed as described in Section~\ref{sec:glx_sample}, however in the following analysis we treat NED and SIMBAD sources separately.
We matched the combined list of NED-only, SIMBAD-only, and NED+SIMBAD sources to \textit{Gaia} sources using the
\textcode{match\_to\_catalog\_sky} method of the \textcode{astropy.coordinates.SkyCoord} class.
Matching was performed individually for each source in the combined list, meaning that it is not strictly one-to-one, and was
based only on sky position.
The positional data compiled in NED and SIMBAD are heterogeneous and generally do not contain uncertainties.
In comparing NED and SIMBAD sources in LVG galaxies, \citet{ref:comparing_NS}
experimented with different matching tolerance values and found a rapid increase in the number of matches 
with tolerance up to 2~arcsec, and little increase beyond tolerances of 5~arcsec. 
We adopt 5~arcsec tolerance for our matching: 
while the \textit{Gaia} astrometric precision and accuracy are both much smaller than this,
using a generous matching radius accounts for the literature uncertainties.

For sources appearing in both NED and SIMBAD, both coordinates were matched to \textit{Gaia}. 
If the NED and SIMBAD coordinates for the same source matched to different \textit{Gaia} sources, we selected as the correct \textit{Gaia} match the source that was nearer to (but within 5~arcsec of) one of the NED or SIMBAD coordinates.
Although \textit{Gaia} DR2 and DR3 identifiers are provided for some SIMBAD sources,% 
\footnote{see \url{https://cds.u-strasbg.fr/gaia}}
we emphasize that \emph{these identifiers were not used in our matching process} (see also Section~\ref{sec:matches}.
The criteria used by SIMBAD to perform that matching were relatively strict and we chose a more relaxed positional match criterion to encompass possible astrometric uncertainties in the NED and SIMBAD source material.

\subsection{Comparing source classifications between databases}
\label{sec:comparing_classifications}

A challenge when comparing classifications of astronomical sources is that there is no completely unambiguous 'ground truth': the measurements upon which classification is based are necessarily limited and uncertain.
\citet{ref:comparing_NS} compared NED and SIMBAD classifications for a very similar dataset to that analysed here --- 75164 sources projected within the Holmberg radius of galaxies in the LVG  --- and found 88 per cent agreement between their classifications, despite very different classification schemes.
From this we conclude that NED and SIMBAD classifications are reasonably robust and using them the basis for comparison to \textit{Gaia} classification is justified.
Because NED contains about ten times as many sources as SIMBAD near LVG galaxies \citep[see Fig.~4 of][and \autoref{fig:venn_diagram} of this work]{ref:comparing_NS}, and also because the NED and SIMBAD classifications do have some disagreement, we compare \textit{Gaia} classifications to NED and SIMBAD separately, rather than constructing a combined `literature' 
classification.

Detailed information on the classes used by NED and SIMBAD is provided in Appendix~\ref{appendix-class}.
For each source type in the NED and SIMBAD classification schemes, we assigned a \textit{Gaia} analogue classification as either one of five classes (star, galaxy, quasar, physical binary star, and white dwarf), or unclassified.
This last category was used for sources found within galaxies such as supernova remnants or molecular clouds, as well as transient events and other phenomena not well captured by the \textit{Gaia} scheme.
Our assignments are given in \autoref{table:ned-gaia} and \autoref{table:simbad-gaia}.
For sources classified by SIMBAD as a candidate of a source type, we
considered it to be of that type (e.g. a SIMBAD `galaxy candidate' was considered to be a galaxy)
in the main comparison. We also considered \textit{Gaia} classification of the SIMBAD candidates separately.
The majority of NED sources (1.01M/1.31M) in nearby galaxy fields are classified only by emission wavelength (e.g. ``ultraviolet excess source'', ``centimetric radio source''); a small number of SIMBAD sources also have such an ambiguous classification.
The final rows of \autoref{table:ned-gaia} and \autoref{table:simbad-gaia}
list these ambiguous source types in NED and SIMBAD, respectively, and we consider them separately (Section~\ref{sec:ambig}).
Table~\ref{table:src_counts} summarises the down-selection and matching process: the parent sample for the following analysis involves 404173 \textit{Gaia} sources in 1040 nearby galaxies.

\section{Results}
\label{sec:results}

\subsection{Matches between databases}
\label{sec:matches}

\begin{figure}
    \centering
    \includegraphics[width=0.4\textwidth]{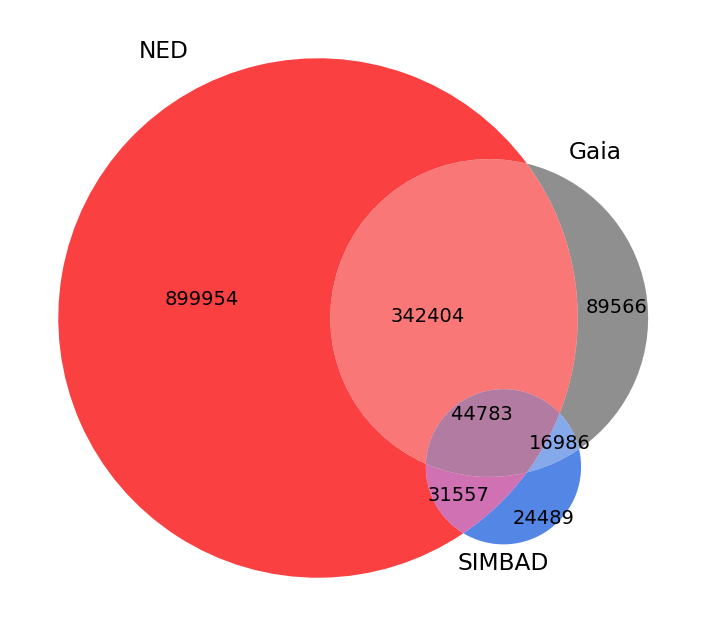}
    \caption{Overlap between 493739 \textit{Gaia}, 1318698 NED and 117815 SIMBAD sources in the vicinity of 1401 nearby galaxies from the Local Volume Galaxy catalogue, matched at 5~arcsec tolerance.} 
    \label{fig:venn_diagram}%
\end{figure}

The numbers of NED, SIMBAD, and \textit{Gaia} sources in nearby galaxies are discussed in detail elsewhere \citep{ref:comparing_NS,ref:gaia_foregrounds}.
Our results are not expected to be identical to those works, due to the evolving nature of both the parent LVG sample and the literature compilations in NED and SIMBAD, but are broadly consistent.
Over 1285 galaxies, we found 1360173 sources in NED and SIMBAD: 1318698 in NED, 117815 in SIMBAD, and 76340 sources appearing in both databases.
About 29~per~cent of NED sources, 52~per~cent of SIMBAD sources, and 59~per~cent of NED+SIMBAD sources were matched to \textit{Gaia} sources, with an overall match fraction of 30~per~cent, comprising 324922 unique \textit{Gaia} matches for 404173 sources in 1040 galaxies.
Approximately 80~per~cent of \textit{Gaia} sources in the vicinity of a nearby galaxy were matched to a NED and/or SIMBAD source, with
the overlaps between NED, SIMBAD, and \textit{Gaia} sources in nearby galaxies illustrated in \autoref{fig:venn_diagram}. 
Only a small fraction (about 3 per cent) of the 61769 SIMBAD/\textit{Gaia} matches had GDR3 IDs in the SIMBAD \textcode{main\_id} field, validating our choice to use a larger matching radius than was used by SIMBAD for \textit{Gaia} matching.

\begin{figure}%
    \centering
    \includegraphics[width=0.45\textwidth]{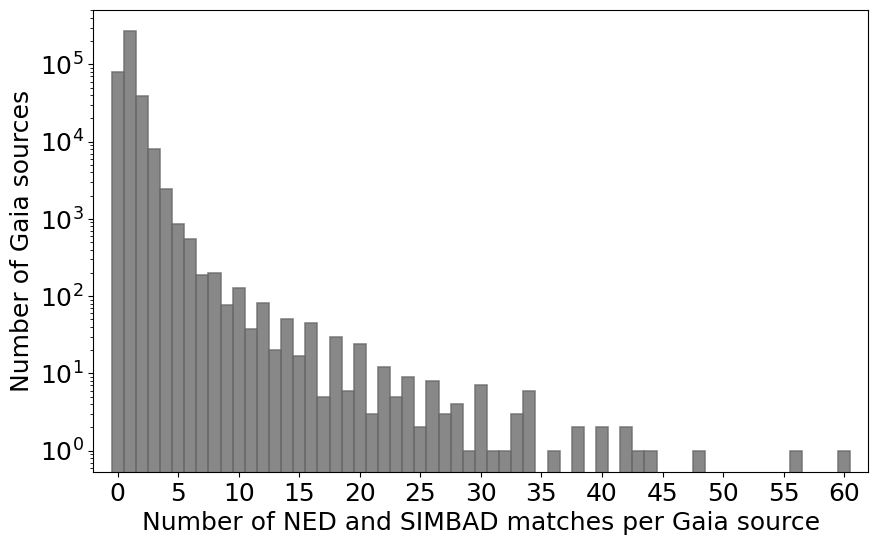}%
    \caption{Distribution of the NED/SIMBAD matches per \textit{Gaia} source (404173 total matches). 
    The peak at one indicates that the majority of sources in NED and SIMBAD have one \textit{Gaia} analogue within a 5\farcs0 match radius.}
    \label{fig:matches_histogram}%
\end{figure}

Coordinate-based matching between NED, SIMBAD, and  \textit{Gaia} sources in the vicinity of nearby galaxies generally works well.
\autoref{fig:matches_histogram} shows the distribution of the number of NED and SIMBAD matches per \textit{Gaia} source;
the mean number of matches per \textit{Gaia} source is 1.0 and the maximum number of matches for a \textit{Gaia} source is 64. 
Most matched \textit{Gaia} sources (84~per~cent) have a single NED or SIMBAD match, with
12~per~cent having two matches, and the remaining 4~per~cent three or more matches. 
Some extreme multiple-match cases are examined in more detail in Section~\ref{sec:Discussion}.
As a conservative approach for the analysis that follows, we removed all \textit{Gaia} sources that are matched to more than one NED and/or SIMBAD source from the comparison. This reduces the number of sources involved to 272422.

\begin{figure}%
    \centering
    \includegraphics[width=0.45\textwidth]{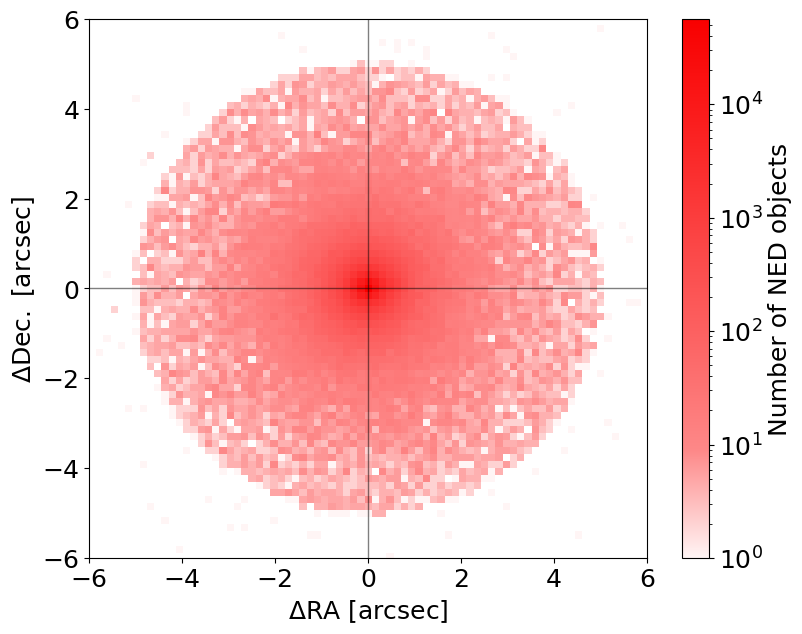}%
    
    \includegraphics[width=0.45\textwidth]{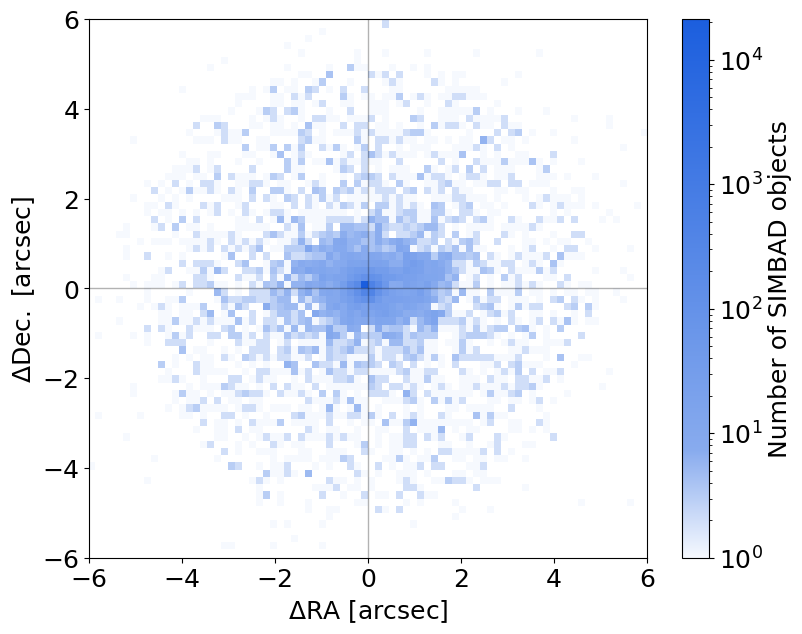}
    \caption{Top: Two-dimensional histograms of coordinate offsets between 265431 NED sources and their corresponding \textit{Gaia} matches. 
     Bottom: Two-dimensional histograms of coordinate offsets between 30172 SIMBAD sources and their corresponding \textit{Gaia} matches.
    Both distributions are centred on zero offset, as expected.
    }
    \label{fig:separation}%
\end{figure}

\begin{figure}%
    \centering
    \includegraphics[width=0.45\textwidth]{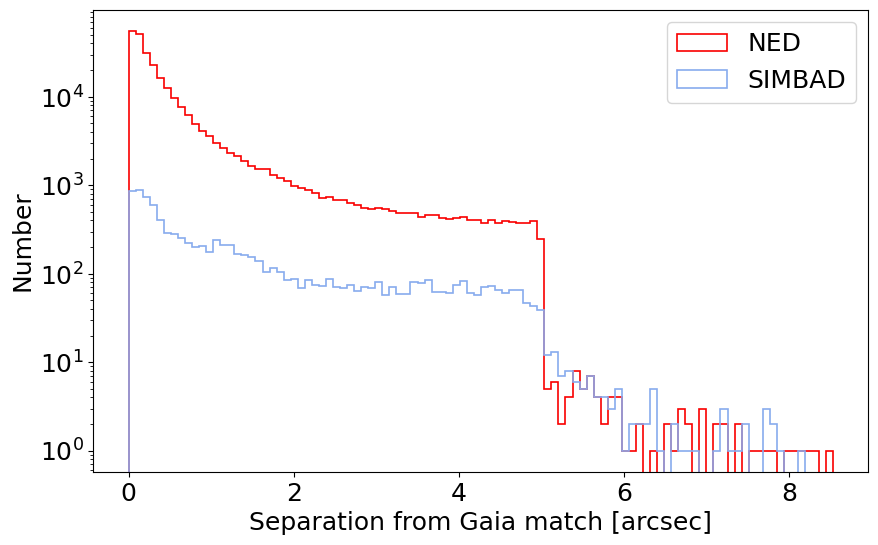}%
    \caption{Distribution of coordinate offsets between 265431 (30172) NED (SIMBAD) sources and their corresponding \textit{Gaia} matches.} 
    \label{fig:match_sep}%
\end{figure}

The coordinate offsets between NED and SIMBAD sources and their corresponding \textit{Gaia} matches are plotted in \autoref{fig:separation}. 
Mean offsets in each of right ascension and declination are $<0\farcs02$ for both NED and SIMBAD.
The few sources outside the $5\farcs0$ boundary visible in \autoref{fig:separation} are the
result of situations where NED and SIMBAD coordinates for the same source differed enough that (e.g.) the
NED coordinates for the \textit{Gaia} match were within 5~arcsec and the SIMBAD coordinates were not.
Figure~\ref{fig:match_sep} shows the distribution of separations between NED/SIMBAD sources and their corresponding \textit{Gaia} matches; the average separations are  $0\farcs57$ and $0\farcs44$,
with 90 per cent of separations within $1\farcs49$ and $1\farcs59$, for NED and SIMBAD
respectively. 
Given that \textit{Gaia} astrometric precision is significantly better than these values, this indicates that 
NED and SIMBAD astrometric uncertainties dominate the matching success.

The most common identifiers for \textit{Gaia}-matched, non-ambiguous NED sources are `WISEA' and `SDSS', referring to the AllWISE data release \citep{cutri2013} and Sloan Digital Sky Survey \citep{sdss}.
These two surveys provide about 90 per cent of NED identifiers.
For the \textit{Gaia}-matched, non-ambiguous SIMBAD sources, no single survey provides more than 10 per cent of SIMBAD identifiers. The most common identifiers are the Tycho catalogue and a survey of stellar velocities in Local Group dwarf galaxies by \citet{walker09}.
One important identifier in the SIMBAD sources relates to \textit{Gaia} itself: SIMBAD sources for which  \textit{Gaia} was
the original and only source should not be included in a comparison between the databases.
We identify these sources by their SIMBAD label:
the default SIMBAD label for \textit{Gaia} sources was set  to ``star", and was only updated if the source was confirmed by other surveys (anonymous referee, priv. comm.).
Of the 1827 such SIMBAD sources, about half have SIMBAD class `*' and 
the rest are various sub-classes of star or star candidate (e.g. `RB?'). 
We have filtered out the former but not the latter in our comparison of classifications. 

\subsection{Comparing \textit{Gaia} and literature classifications}

\begin{table*}
\centering
\caption{Classifications of \textit{Gaia} matches to NED and SIMBAD objects}
\label{table:gaia_classifications}
\begin{tabular}{lrrrrr}
 & Gaia/NED & NED/Gaia & Gaia/SIMBAD & SIMBAD/Gaia & Gaia/(NED or SIMBAD)\\
\hline
\hline
Star         &   41692 &   37271 &   25127&   24595&  260037\\
Galaxy       &    2569 &    7740 &    1392&    1630&    4237\\
Quasar       &    2447 &     207 &    2058&     674&    5052\\
Binary star  &      85 &       5 &      85&     356&     289\\
White dwarf  &      85 &       5 &     332&     342&     519\\
Unclassified &     300 &    1950 &     410&    1807&    2288\\
Total        &   47178 &   47178 &   29404&   29404&  272422\\
\hline
\multicolumn{6}{l}{Note: final column (only) includes sources with ambiguous classifications in NED and/or SIMBAD.} 
\end{tabular}
\end{table*}

\autoref{table:gaia_classifications} gives the \textit{Gaia} classifications for sources matched to NED and SIMBAD sources,
where we include only non-ambiguous NED and SIMBAD sources (columns 2--5) and all sources (column 6).
Nearly all of the  NED and SIMBAD sources matched to \textit{Gaia} sources have \textit{Gaia} classifications: only $(2288/272422)=0.8$~per~cent do not.
\textit{Gaia} classifies roughly 95~per~cent of the matched sources as stars,
about 2~per~cent each as galaxies and quasars, $<1$~per~cent as either binary stars or white dwarfs.
This distribution differs substantially from the overall \textit{Gaia} class distribution, which has a higher fraction of stars.
\citet{ref:DR3_source_classification} noted that \textit{Gaia} sources in the vicinity of the Magellanic Clouds include an excess of sources classified as quasars according to \textcode{classprob\_dsc\_combmod} and this appears to be the case for other nearby galaxies as well.

\begin{figure}
    \centering
    \includegraphics[width=0.45\textwidth]{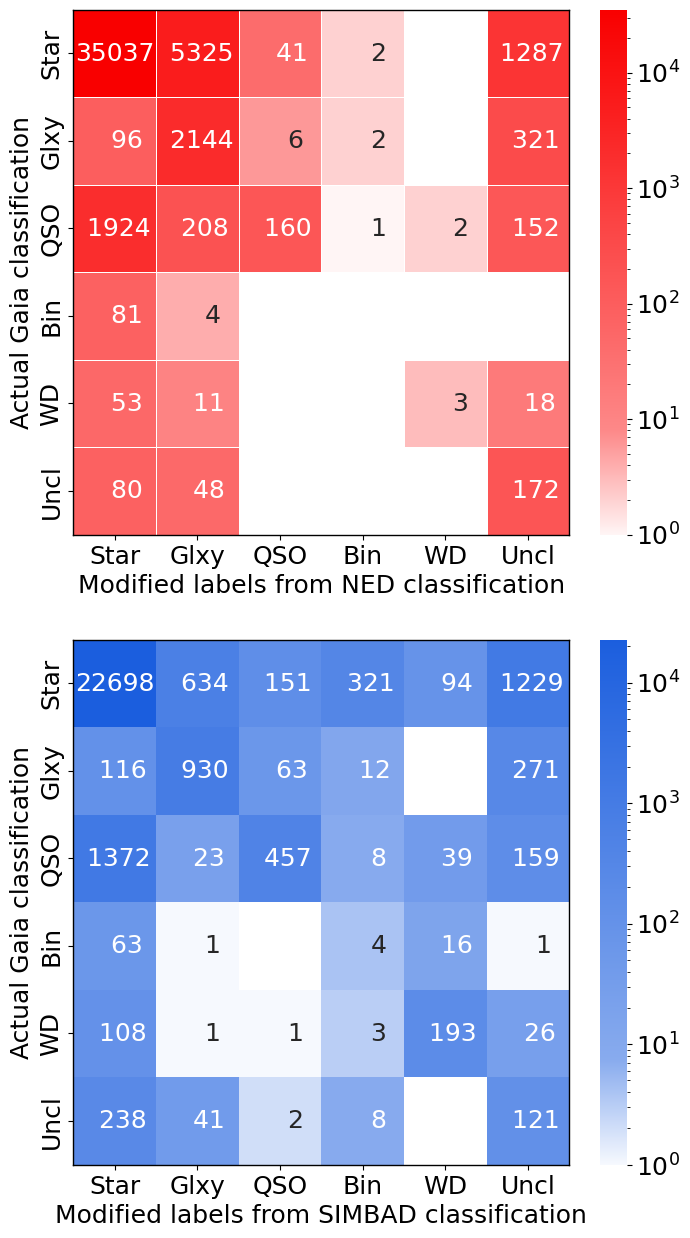}%
    \caption{Expected vs. actual \textit{Gaia} classification based on NED classification (\textit{top}, $N=47178$) and SIMBAD classification (\textit{bottom}, $N=29404$), non-ambiguous sources only.    
    }
    \label{fig:confusion_matrix}%
\end{figure}

We first assess the performance of \textit{Gaia} classification against NED or SIMBAD classifications which are 
unambiguous (e.g. `quasar' or `carbon star' in contrast to `infrared source' or `ultraviolet source'); there are approximately 
47 and 29 thousand such sources for NED and SIMBAD respectively.
This type of analysis generally presumes that one has a test set of sources for which the true classification is known, 
and for which a predictive model is being validated.
Here, as in many cases in astronomy, there is no `ground truth': to compute confusion matrices and performance metrics we assume the NED or SIMBAD classification to be the truth, and the \textit{Gaia} classification to be the prediction.

\autoref{fig:confusion_matrix} shows the confusion matrix between the expected \textit{Gaia} classification based on the source's classification in the literature and the actual \textit{Gaia} classification of the source's \textit{Gaia} match.
The overall pattern is quite similar between NED and SIMBAD:
sources classified as stars by NED or SIMBAD and by \textit{Gaia} make up the majority of sources.
The most commonly confused classes (the off-diagonal entries) are between stars and another class: either literature-classified galaxies which \textit{Gaia} classifies as stars, or literature-classified stars which \textit{Gaia} classifies as quasars.
Sources noted as unclassified in the literature (this includes categories such as `star cluster' and `supernova remnant'; see \autoref{table:ned-gaia} and \autoref{table:simbad-gaia}) are most likely to be classified by \textit{Gaia} as stars (see Section~\ref{sec:unclass}).
Sources for which the classifications agreed between NED and \textit{Gaia} were on average brighter and redder than sources for which they disagreed, with mean $G$ magnitudes and $B_p-R_p$ colours of
18.97 and 1.44 for agreeing sources versus 20.49 and 1.17 for disagreeing sources.  
For SIMBAD the corresponding mean values are $G=17.29$ and $B_p-R_p=1.22$ for agreeing sources and
19.86 and 0.86 for disagreeing sources.

\autoref{table:overall_metrics} summarises the overall performance metrics for \textit{Gaia} classification.
The metric definitions are those used in \textcode{scikit-learn} \citet{scikit-learn} and are given in Appendix~\ref{appendix-metric}.
Intuitively, accuracy is the fraction of classifications that are correct,
balanced accuracy is the fraction of correct classifications computed per class and averaged over classes,
purity (precision) is the fraction of sources in a class that are truly members of that class,
and completeness (recall) is the fraction of sources in a class that are classified as such. 
Computed over all objects, accuracy, purity, and completeness are all 0.80 when NED classes are considered as truth values, and 0.83 when SIMBAD classes are considered as truth values.
The balanced accuracy (averaged over the six classes) is 0.45 for NED classes as truth values, and 0.47 for SIMBAD classes as truth values, with similar completeness values, and lower purity.

\begin{table}
\centering
\caption{Classification metrics: \textit{Gaia} performance versus NED and SIMBAD predictions, non-ambiguous sources only}
\label{table:overall_metrics}
\begin{tabular}{lrr}
\hline
&  NED & SIMBAD\\
&  $N=47178$ & $N=29404$\\
\hline
Global values over all objects & & \\
Purity (precision) & 0.80 & 0.83\\
Completeness (recall) & 0.80 & 0.83\\
Accuracy     & 0.80 & 0.83\\
Averages over per-class values& & \\
Purity (precision) & 0.39 & 0.45\\
Completeness (recall) & 0.45 & 0.47\\
Accuracy -- balanced & 0.45 & 0.47\\
\hline
\end{tabular}
\end{table}

Breaking down the metrics by class enables a more detailed understanding of classification performance.
As shown in \autoref{table:class_metrics}, there is considerable variation in performance between classes. 
\textit{Gaia} performs well (metrics $>0.8$) on classifying single stars as identified by both NED and SIMBAD, although as \citet{ref:DR3_source_classification} note, the class imbalance (most \textit{Gaia} sources are stars) makes this metric not particularly meaningful for this class.
\textit{Gaia} performance on literature-identified galaxies shows purity $>0.6$ and lower completeness (0.3--0.6).
On literature-identified quasars, \textit{Gaia} completeness
(0.7--0.8) is much higher than purity (0.1--0.2), implying that \textit{Gaia} false positives outnumber false negatives: 
the \textit{Gaia} classification performs well at identifying true quasars and less well at rejecting non-quasars. 
The small number of NED-predicted quasars implies that the metrics for this class are subject to significant uncertainty. 
Even more uncertain are the metrics for the small number of NED-identified white dwarfs or binary stars.
\textit{Gaia} purity and completeness are low for SIMBAD-identified binaries (0.05 and 0.01, respectively) and moderate
(0.58 and 0.56, respectively)  for SIMBAD-identified white dwarfs.

\begin{table}
\centering
\caption{Classification metrics: \textit{Gaia} performance versus NED and SIMBAD predictions, per class, non-ambiguous sources only}
\label{table:class_metrics}
\begin{tabular}{l c c  r}
\hline
Class & Precision & Recall &  Support\\
& (purity) & (completeness) & $N$\\
\hline
NED predictions& & &\\
Star         &  0.84 &  0.94  &   37271\\
Galaxy       &  0.83 &  0.28  &    7740\\
Quasar       &  0.07 &  0.77  &     207\\
Binary star  &  0.00 &  0.00  &       5\\
White dwarf  &  0.04 &  0.60  &       5\\
Unclassified &  0.57 &  0.09  &    1950\\
Weighted avg &  0.82 &  0.80  &   47178\\
\hline
SIMBAD predictions& & &\\
Star         &  0.90 &  0.92  &   24595\\
Galaxy       &  0.67 &  0.57  &    1630\\
Quasar       &  0.22 &  0.68  &     674\\
Binary star  &  0.05 &  0.01  &     356\\
White dwarf  &  0.58 &  0.56  &     342\\
Unclassified &  0.30 &  0.07  &    1807\\
Weighted avg &  0.82 &  0.83  &   29404\\
\hline
\end{tabular}
\end{table}

\begin{table}
\centering
\caption{Classification metrics: \textit{Gaia} performance versus NED and SIMBAD predictions, per class, non-ambiguous sources only}
\label{table:class_metrics_pure}
\begin{tabular}{l c c  r}
\hline
Class & Precision & Recall &  Support\\
& (purity) & (completeness) & $N$\\
\hline
NED predictions& & &\\
Star         &  0.00 &  0.00  &     272\\
Galaxy       &  0.93 &  0.83  &     224\\
Quasar       &  0.19 &  0.99  &      75\\
Unclassified       &  0.00 &  0.00  &      21\\
Weighted avg &  0.38 &  0.44  &     592\\
\hline
SIMBAD predictions& & &\\
Star         &  0.00 &  0.00  &     205\\
Galaxy       &  0.71 &  0.98  &     101\\
Quasar       &  0.45 &  0.92  &     191\\
White dwarf           &  0.00 &  0.00  &       6\\
Unclassified      &  0.00 &  0.00  &      25\\
Weighted avg &  0.30 &  0.52  &     528\\
\hline
\end{tabular}
\end{table}

We can also compare performance of the \textit{Gaia} `pure' samples of galaxies and quasars \citep{ref:DR3_extragalactic_content} against NED and SIMBAD classifications.
Information on the pure samples is not directly available from the main DR3 data tables; we retrieved the \textcode{gaiadr3.astrophysical\_parameters} information
for the \textit{Gaia} sources matched to NED and/or SIMBAD source and recreated the \textcode{classlabel\_dsc\_joint} parameter as described by \citet{ulla2022}.
Most of the matched \textit{Gaia} sources had \textcode{classlabel\_dsc\_joint} as `unclassified': 751 were in the pure quasar sample and 352 in the pure galaxy sample. 
Of these 1103 objects, 592 (528) had non-ambiguous NED (SIMBAD) classifications.
Computing classification metrics per class  between the pure and non-pure \textit{Gaia} samples (see \autoref{table:class_metrics_pure}), we find that completeness and purity are higher for the pure sample, for both galaxies and quasars. 
The comparative results apply to both NED and SIMBAD classifications.
The most common classification mismatches were \textit{Gaia} galaxies classified as star clusters, \ion{H}{ii} regions, or molecular clouds by NED or SIMBAD (see Section~\ref{sec:unclass}) and \textit{Gaia} quasars classified as stars by NED or SIMBAD.

As mentioned in Section~\ref{sec:comparing_classifications}, we included SIMBAD `candidates' as equivalent to confirmed classifications in the previous analysis. 
We also considered sources with SIMBAD `candidate' classifications
separately to see how they affected overall performance.
In the vicinity of nearby galaxies, there are 12128 sources classified as a candidate of a type in SIMBAD, of which 3461 have a \textit{Gaia} match within 5 arcseconds.
The overall accuracy of \textit{Gaia} classification is 0.88.
The vast majority (2850/3461) of the SIMBAD candidate sources  are candidate stars, which \textit{Gaia} classifies as stars with purity 0.95 and completeness 0.93. 
The next most populated category ($N=306$) is white dwarf candidates, which \textit{Gaia} classifies as white dwarfs with purity 0.98 and completeness 0.58.
In both cases the metric values are higher than for the full set of SIMBAD sources.
The remaining categories have $<100$ sources: except for completeness of 0.67 on $N=30$ quasars,
all other per-class metrics are $<0.3$.
We conclude that SIMBAD  `candidate' classifications are at least as reliable a source of ground truth as non-candidate classifications.

\subsection{`Unclassified' sources}
\label{sec:unclass}

\begin{figure}%
    \centering
    \includegraphics[width=0.45\textwidth]{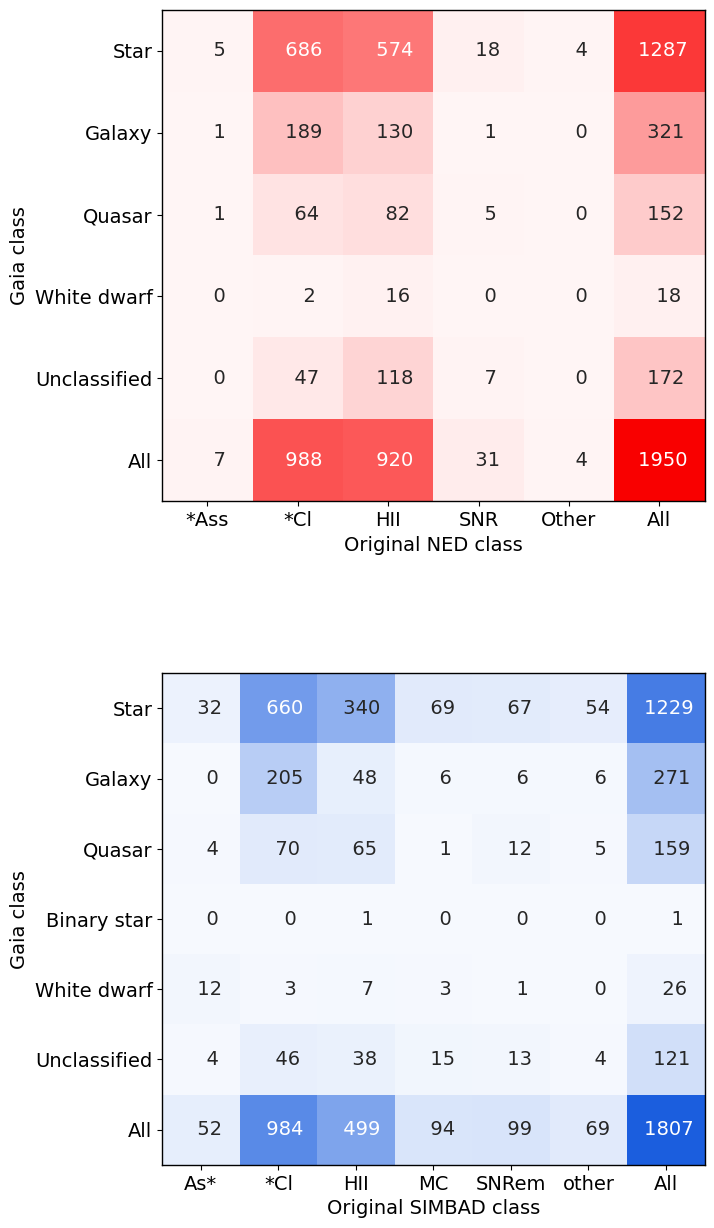}
    \caption{\textit{Gaia} classification versus NED  (\textit{top}, $N=1950$) and SIMBAD (\textit{bottom},$N=1807$) original classification, for NED and SIMBAD `unclassified' sources (classes that did not map onto the \textit{Gaia} categories).
For NED, `Other' replicates the NED category of the same name.
    For SIMBAD, `Other' includes SIMBAD's unknown category (symbol `?') and other categories such as HVC, reg, and ev.}
    \label{fig:unclass}%
\end{figure}

The right-hand columns of the confusion matrices in Figure~\ref{fig:confusion_matrix} represent sources whose
NED or SIMBAD classifications did not map onto the five \textit{Gaia} categories (see the corresponding rows of Tables~\ref{table:ned-gaia} and \ref{table:simbad-gaia}).
Except for the SIMBAD `unknown' and `error' categories, these sources are not truly `unclassified' (but see the following section on ambiguous literature classifications). 
Figure~\ref{fig:unclass} shows the breakdown of the literature classifications for these sources and their \textit{Gaia} classifications. 
The 1950 NED-unclassified sources are primarily star clusters, \ion{H}{ii} regions, and molecular clouds; their \textit{Gaia} classifications are 66 per cent star, 16 per cent galaxy, 8 per cent quasar, 1 per cent white dwarf, and 9 per cent unclassified.
The 1807 SIMBAD-unclassified sources fall into the same major categories, although in different proportions, as well as many smaller categories.
Their \textit{Gaia} classifications are similar to NED: 68 per cent star, 15 per cent galaxy, 9 per cent quasar, 1 per cent white dwarf, and 8 per cent unclassified.
It is unsurprising that the  \textit{Gaia} classification distribution is similar, as there are 736 sources in common between the
NED-unclassified and SIMBAD-unclassified sources.

The bottom rows of the confusion matrices in Figure~\ref{fig:confusion_matrix} represent sources unclassified by \textit{Gaia}.
While one might conjecture that these sources are those for which the \textit{Gaia} classifier was undecided (i.e. no class probabilities $>0.5$; see Section~\ref{sec:gaia_classifications}), Upon further investigation we found that this was not the case.
Most of these sources (283 of 300  \textit{Gaia}-unclassified matches to NED,
and 395 of 410 \textit{Gaia}-unclassified matches to SIMBAD) were missing \textit{Gaia} astrometric or photometric measurements, such that \emph{all} class probabilities were undefined.

The detailed literature classifications allow us to further investigate the nature of \textit{Gaia}-unclassified sources.
Figure~\ref{fig:unclass} shows that the NED/\textit{Gaia} `unclassified' sources are primarily \ion{H}{ii} regions, star clusters, and supernova remnants; the
SIMBAD-`unclassified' source fall into the same categories although with different proportions.
Not shown in Figure~\ref{fig:unclass} are the 80 NED `stars' (62 stars, 14 variable stars, 3 planetary nebulae, 1 Wolf-Rayet star) matched to \textit{Gaia}-unclassified sources  or the 48 NED `galaxies' (40 galaxies, 6 `parts of galaxies',%
\footnote{Refers to sources within a galaxy whose physical nature was unspecified
(e.g. a resolved source that could be a nebula or star cluster) or that did not fit into any existing classes (e.g. tidal disruption events; M. Schmitz, priv. communication.)
}
2 galaxy groups) matched to \textit{Gaia}-unclassified sources.
SIMBAD `stars', `galaxies', and `quasars' matched to \textit{Gaia}-unclassified sources are SIMBAD-classified into many sub-categories with only a few members each.
We can infer that the \textit{Gaia} `failure' to classify many of these sources is consistent with their literature-identified nature: 
172 of 300 NED sources and 121 of 410 SIMBAD sources are literature-classified as categories other than stars, galaxies, or quasars.
However, as shown below,  \textit{Gaia}-unclassified sources cannot be used to compile a complete sample of such galaxy components, since many do have  \textit{Gaia} classifications.

\subsection{Ambiguous literature classifications}
\label{sec:ambig}
\begin{figure}%
    \centering
    \includegraphics[width=0.45\textwidth]{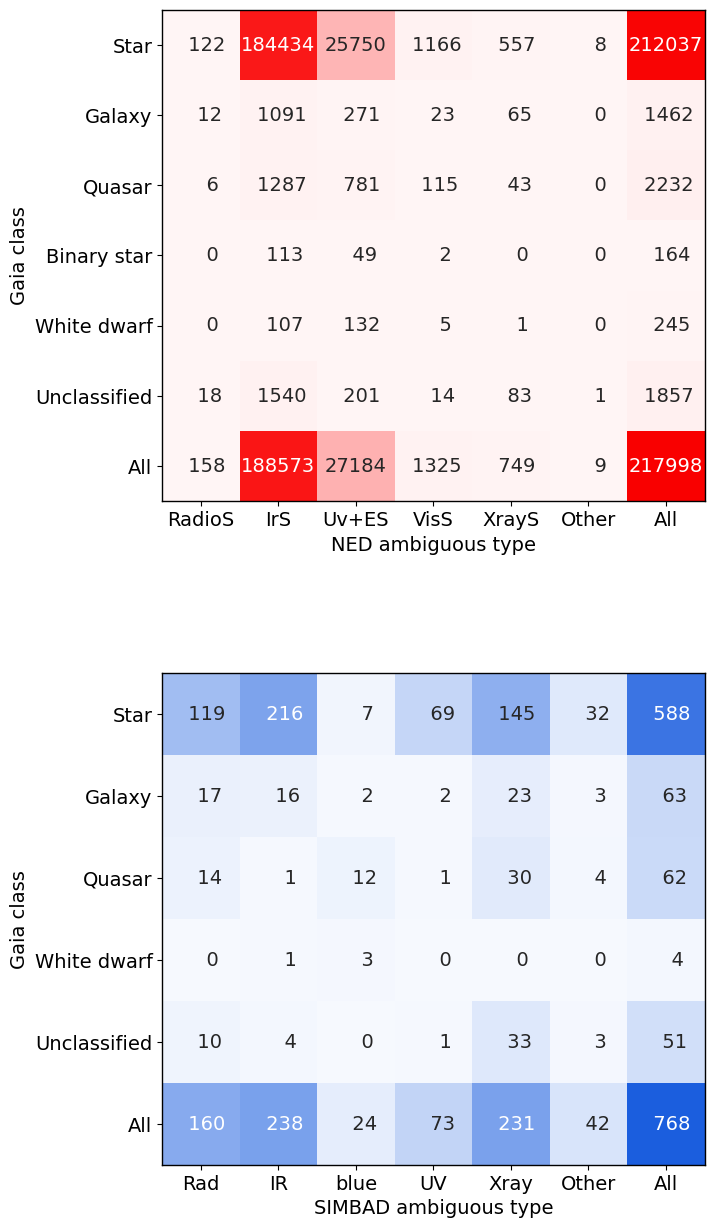}
    \caption{\textit{Gaia} classification versus NED  (\textit{top}, $N=217998$) and SIMBAD (\textit{bottom}, $N=768$) original classification, for NED and SIMBAD `ambiguous' sources (classes that did not map onto the \textit{Gaia} categories).
    For NED, `Uv+ES' includes ultraviolet sources (UvS) and ultraviolet excess sources (UvES) and `Other' includes emission-line sources (EmLS, EmObj) and gamma-ray sources (GammaS).
    For SIMBAD, `IR' and `Radio' are sums of subcategories (e.g. far/mid/nearIR, cm/mm/smmRad)
    and `Other' includes emission-line sources (EmObj), gamma-ray sources (gamma) and gamma-ray bursts (gammaBurst).
    }
    \label{fig:ambig}%
\end{figure}

Finally, we examine the use of \textit{Gaia} to classify sources with ambiguous classifications (see final rows of Tables~\ref{table:ned-gaia} and \ref{table:simbad-gaia}) in the literature.
In the vicinity of nearby galaxies, we find 217998 sources with ambiguous classifications in NED and 768 sources with ambiguous classifications in SIMBAD, of which about one third each of NED and SIMBAD sources have a \textit{Gaia} match within a 5~arcsec radius.
Figure~\ref{fig:ambig} shows the distribution of these sources across the \textit{Gaia} labels.
Most of the NED sources are infrared or ultraviolet sources classified by \textit{Gaia} as stars (85 and 12~per~cent, respectively).
In the much smaller number of ambiguous SIMBAD sources, the largest classes are
near-infrared, X--ray, and radio sources classified as stars (26, 18 and 13~per~cent, respectively), and X--ray and infrared sources classified as galaxies (3 and 2~per~cent)  or quasars (4 and 2~per~cent).
About 7~per~cent of the ambiguous SIMBAD sources are unclassified by \textit{Gaia}.

\section{Discussion}
\label{sec:Discussion}

One limitation of this work is the case of multiple NED and SIMBAD sources matched to the same \textit{Gaia} source.
While \textit{Gaia}'s astrometry is, as expected, excellent (\autoref{fig:separation}),
the limitations of its spatial resolution and/or the literature astrometry are apparent in the number of matches per source (\autoref{fig:matches_histogram}).
The multiple matches
typically correspond to sources classified by NED and SIMBAD as located in star clusters or molecular clouds, sources in small and distant galaxies, or sources in regions where two or more galaxies overlap. 
The greatest number of NED and SIMBAD matches for a single \textit{Gaia} source was shared by two \textit{Gaia} sources. 
The first of these corresponds to a \textit{Gaia} source matched to multiple star clusters and molecular clouds in the crowded central region of the interacting galaxy M51.
The star cluster catalogue was constructed from Hubble Space Telescope imaging \citep{ref:M51_2008} with significantly higher spatial resolution than \textit{Gaia}.
The millimetre-wave observations used to make the catalogue of molecular clouds \citep{ref:M51_2014} had coarser spatial resolution (beam approximately 1\arcsec) but detected sources at a much higher spatial density than \textit{Gaia} sources in the region.
The second example of many NED/SIMBAD source matching a single \textit{Gaia} corresponds to observations of a molecular cloud in the dwarf galaxy ESO495-021 by \citet{ref:ESO495-021_2018} and \citet{ref:ESO495-021_2019}.
Again these observations had high spatial resolution ($\sim 0\farcs3$) and a high spatial density of sources compared to \textit{Gaia} sources in the region. 

The multiple matches described above are examples of the inherent complexity in cross-matching observations made in different wavelength regimes and with different spatial resolution.
The parent LVG catalog is an all-sky survey, and crowding and classification errors in the Galactic plane \citep[see, e.g.,][]{marchantcortes2024} are one facet of this complexity. 
However, as only $\sim5$ percent of \textit{Gaia} sources matched with NED or SIMBAD sources are at Galactic latitudes $<10\deg$, we do not expect proximity to the Galactic plane to be a major factor in our comparison.
In this work, we made the choice to simply remove any multiply-matching \textit{Gaia} sources
from the classification comparison.
More sophisticated cross-matching techniques, such as probabilistic methods 
\citep[e.g.][]{pineau2017} or the use of Bayesian priors on colour and/or magnitude \citep[e.g.][]{salvato2018} could potentially resolve ambiguity and yield a more comprehensive set of one-to-one matches.

Another limitation is the lack of a `ground truth' classification, including the large fraction of literature sources with ambiguous classifications.
Although the good agreement between NED and SIMBAD classifications found by \citet{ref:comparing_NS} is encouraging, in many cases those classifications are drawn from the same original literature source, so it would be surprising if they did not agree.
We did not exclude from our analysis sources for which the two literature databases' classification disagreed \citep[12 per cent in the results of][]{ref:comparing_NS}.
A possible avenue for future work would be to use \textit{Gaia} classifications and measurement to further investigate sources with conflicting classifications, as in Section~\ref{sec:ambig} for ambiguous sources.

In general, correctly distinguishing between foreground, background, and in-galaxy sources projected near nearby galaxies is a challenging task: for example, catalogues of globular cluster candidates in the Andromeda galaxy include large numbers of stars and galaxies \citep{wang2023}. 
When compared to literature classifications, the \textit{Gaia} classifications for background galaxies in the vicinity of nearby galaxies have moderate purity (0.83, 0.67 for NED and SIMBAD respectively) and completeness (0.28, 0.57).
Background quasars had  lower purity (0.07 and 0.22 for NED and SIMBAD respectively) but higher completeness (0.77, 0.68).
Restricting the comparison to the `purer' \textit{Gaia} classifications improves completeness and purity for both galaxies and quasars.
This is consistent with the results of \citet{ref:DR3_source_classification} and suggests that \textit{Gaia} measurements in these crowded fields (that may also be affected by extinction within the nearby galaxy) must be treated carefully.
It may be possible to improve the training for \textit{Gaia} classifications in these environments; for example,
the training set for \textit{Gaia's} `star' class is comprised of arbitrarily chosen \textit{Gaia} DR3 sources that were not spectroscopically confirmed to be a galaxy or quasar \citep{ref:DR3_source_classification}, meaning that this classification potentially encompasses a wide variety of point sources. 
Catalogues of sources in the \textit{Gaia} DR3 with confirmed classifications in the literature, such as the catalogue of variable sources compiled by \citet{ref:lit_DR3_variable_objects}, could be used as training sets for new classes.

The present work shows that broad \textit{Gaia} classifications from its Discrete Source Classifier do not map particularly well onto the nature of sources near nearby galaxies, as given in the literature, although the small sample of galaxies and quasars in the purer \textit{Gaia} sample does have higher classification metrics.
A mismatch between \textit{Gaia} DSC classifications and the literature classes is not unexpected, given that the vicinity of nearby galaxies is not the situation for which DSC was designed.
Previous work \citep{ref:gaia_foregrounds} showed that approximately two thirds of \textit{Gaia} sources in the vicinity of nearby galaxies have astrometric properties consistent with being foreground sources in the Milky Way. The spatial 
distribution of these sources implies that the true foreground fraction may be as high as 90 per~cent. 
Finding the \textit{Gaia} sources physically associated with nearby galaxies will likely need to make use of
many measurements contained in the \textit{Gaia} database (e.g., galactocentric position, colour, variability, image size);
and exploring these is a promising avenue for future research \citep[e.g., see][who used \textit{Gaia} DR3 to search for previously-unidentified globular clusters in M31]{wang2023}.

\section{Summary and conclusions}
\label{sec:Conclusions}

NED, SIMBAD, and \textit{Gaia} DR3 provide different but complementary views of the constituents of nearby galaxies. 
In the directions of the $~\sim1400$ galaxies in the Local Volume Galaxy survey (excluding a few Local Group galaxies),
there are approximately $1.3\times10^6$ NED sources,  $2\times10^5$  SIMBAD sources, and $5.7\times10^5$ \textit{Gaia} sources. 
Matching between these sources gives about $3.2\times10^5$ unique \textit{Gaia} matches for $4\times10^5$ sources over 1040 galaxies.
Most matched \textit{Gaia} sources (84~per~cent) have a single NED or SIMBAD match, and these matched sources are separated from their corresponding \textit{Gaia} matches by approximately 0.5~arcsec, on average.

The classification of \textit{Gaia} sources matched to NED or SIMBAD sources in the vicinity of nearby galaxies
differs substantially from the overall \textit{Gaia} class distribution.  
The NED/SIMBAD-matched sources are classified by  \textit{Gaia} as about 95~per~cent stars,
2~per~cent each galaxies and quasars, and $<1$~per~cent each binary stars or white dwarfs.
Using literature classifications for approximately 47k (29k) NED (SIMBAD) sources with unambiguous (non-`wavelength source') classifications as the `truth' values, \autoref{table:overall_metrics}  shows that \textit{Gaia} classification has accuracy of 0.80 (0.83) compared to NED (SIMBAD).
Classification agreement is better for brighter sources, as might be expected if these sources have been better-studied in the literature and/or have more secure \textit{Gaia} measurements.
\autoref{table:class_metrics} shows that
\textit{Gaia} performance is best (purity and completeness 0.84--0.94) on classifying single stars as identified by both NED and SIMBAD, and decreases (purity  $0.7-0.8$, completeness $0.3-0.6$) for literature-identified background galaxies; these metrics are higher for the purer sample of \textit{Gaia} classifications.
\textit{Gaia} purity is relatively low ($<0.3$), with higher completeness $>0.7$ for literature-identified quasars, particularly those identified by NED; this metric again improves for the purer sample of \textit{Gaia} classifications.
Considering the 3461 Gaia/SIMBAD `candidate' sources classifications separately, we find that \textit{Gaia} classification for these sources has an accuracy of 0.88.
Most are candidate stars, classified as such with purity and completeness $>0.9$; the much smaller number of white dwarf candidates are classified as such with purity of 0.98.

Our analysis also considered sources which did not have a \textit{Gaia} classification, 
whose literature classifications were not well-captured by the \textcode{classprob\_dsc\_combmod} categories (NED and SIMBAD `unclassified', e.g., star clusters, molecular clouds), or whose literature classifications were ambiguous (e.g. `infrared source').
Most sources unclassified by \textit{Gaia} had no class probabilities due to missing photometric or astrometric data,
and a majority had literature classifications that also fell into the `unclassified' category.
The 1950 (1807) NED (SIMBAD) `unclassified' sources which were classified by \textit{Gaia} were distributed across the  
\textit{Gaia} categories (see \autoref{fig:unclass}), with no clear preferences, for example for star clusters to be \textit{Gaia}-identified as galaxies or \ion{H}{ii} regions  to be \textit{Gaia}-identified as quasars.
Most of the 218k ambiguous NED sources matched to \textit{Gaia} sources are infrared or ultraviolet sources classified by \textit{Gaia} as stars (85 and 12~per~cent, respectively).
The 768 ambiguous SIMBAD sources matched to \textit{Gaia} sources include specific-wavelength sources classified as stars (57~per~cent),  galaxies (5~per~cent) or quasars (6~per~cent), or unclassified (7~per~cent). 

We conclude that \textit{Gaia} sources in the vicinity of nearby galaxies differ in their classification distribution   from \textit{Gaia} sources in the (Galactic) field.
Agreement between \textit{Gaia} classification and literature classification found in NED and SIMBAD is best for stars, and decreases for quasars, (background) galaxies and white dwarfs, being lowest for binary stars.
Although many \textit{Gaia} sources in the vicinity of nearby galaxies do not fall into the above five categories, the \textit{Gaia} `unclassified' status does not capture all types of galaxy constituents that are also not well-described by these categories with high completeness.
Nearby galaxies contain a wide variety of objects, from individual stars to star clusters to molecular clouds and supernova remnants, and many of these objects are present in \textit{Gaia} DR3.
Combining NED and SIMBAD classifications into a unified, detailed classification scheme,
and then using sources labelled by NED and SIMBAD as a training set for classifiers based on \textit{Gaia} observations, 
is a potential way forward in unlocking the rich potential of \textit{Gaia} in understanding the contents of nearby galaxies.

\section*{Acknowledgements}

The authors thank the anonymous referee for comprehensive reports which helped to improve the manuscript.
We gratefully acknowledge the use of TOPCAT \citep{ref:TOPCAT}, AstroPy \citep{astropy2018}, pandas \citep{pandas}, 
and scikit-learn \citep{scikit-learn} as well as
support from an NSERC Undergraduate Summer Research Award to JH and an NSERC Discovery Grant to PB.
This work has made use of data from the European Space Agency (ESA) mission
{\it Gaia} (\url{https://www.cosmos.esa.int/gaia}), processed by the {\it Gaia}
Data Processing and Analysis Consortium (DPAC,
\url{https://www.cosmos.esa.int/web/gaia/dpac/consortium}). Funding for the DPAC
has been provided by national institutions, in particular the institutions
participating in the {\it Gaia} Multilateral Agreement.
The Gaia archive website is \href{https://archives.esac.esa.int/gaia}{https://archives.esac.esa.int/gaia}. 
This research has made use of the NASA/IPAC Extragalactic Database (NED),
which is operated by the Jet Propulsion Laboratory, California Institute of Technology,
under contract with the National Aeronautics and Space Administration,  
the SIMBAD database, operated at CDS, Strasbourg, France, 
and the Local Volume Galaxy catalogue maintained by the Special Astrophysical Observatory of the Russian Academy of Sciences.
This research was carried out on the traditional lands of the Anishinaabek, Haudenosaunee, Lūnaapéewak and Chonnonton Nations, on lands connected with the London Township and Sombra Treaties of 1796 and the Dish with One Spoon Covenant Wampum. This land continues to be home to diverse Indigenous Peoples (First Nations, Métis and Inuit) whom we recognize as contemporary stewards of the land and vital contributors of our society.

%%%%%%%%%%%%%%%%%%%%%%%%%%%%%%%%%%%%%%%%%%%%%%%%%%
\section*{Data Availability}

The data underlying this article are available in Zenodo, via \url{https://doi.org/10.5281/zenodo.7767603}. 
The datasets were derived from sources in the public domain: NED (\href{https://ned.ipac.caltech.edu/}{https://ned.ipac.caltech.edu/}), SIMBAD (\href{https://simbad.u-strasbg.fr/simbad/}{https://simbad.u-strasbg.fr/simbad/}), \textit{Gaia} Data Release 3 (\href{https://gea.esac.esa.int/archive/}{https://gea.esac.esa.int/archive/}), and the Local Volume Galaxy catalogue (\href{http://www.sao.ru/lv/lvgdb/}{http://www.sao.ru/lv/lvgdb/}).

\bibliographystyle{mnras}
\bibliography{references}

\appendix
\section{Analogous and ambiguous classifications in NED, SIMBAD and \textit{Gaia}}
\label{appendix-class}

A complete list of NED  types is found at \url{https://ned.ipac.caltech.edu/help/ui/nearposn-list_objecttypes}.
A "!" in the type code indicates a source within the Milky Way; in this work we do not distinguish between Galactic and extragalactic sources of the same type.
We follow the SIMBAD classification hierarchy in (e.g.) distinguishing between `star' and `binary star'; NED
assignments are set up to match this as closely as possible.

\begin{table*}
    \centering
    \caption{NED source types and analogous Gaia classifications}
    \label{table:ned-gaia}
    \begin{tabular}{ll}
    \hline
    Modified Gaia label & NED codes \\
    \hline
    Star & *, Blue*, C*, exG*, Flare*, PN, Red*, SN, V*, WR*, !*, !Blue*, !C*, !Flar*, !PN, !Red*, !SN, !V*, !WR*\\
Galaxy & G, GClstr, GGroup, GPair, GTrpl, G\_Lens, PofG\\
Quasar &  QGroup, QSO, Q\_Lens\\
White dwarf & WD*, !WD*\\
Binary star & **, Nova, !**, !Nova\\
Unclassified &  *Ass, *Cl, AbLS, HII, MCld, Neb, Other, Psr, RfN, SNR, XrayS, EmLS, EmObj, GammaS, IrS, RadioS, UvES, UvS, VisS,\\
    & !*Ass, !*Cl, !EmObj, !HII, !MCld, !Neb, !Psr, !RfN, !SNR\\
Ambiguous & EmLS, !EmObj, RadioS, IrS, VisS, UvS, UvES, XrayS, GammaS\\
\hline
\end{tabular}
\end{table*}

A complete list of SIMBAD types is found at \url{http://simbad.cds.unistra.fr/guide/otypes.htx}.
Table~\ref{table:simbad-gaia} does not include SIMBAD `candidate' classifications (e.g. `galaxy candidate'); these are included in the same class as the corresponding non-candidate.
\begin{table*}
\caption{SIMBAD source types and analogous Gaia classifications}
    \label{table:simbad-gaia}
    \centering
    \begin{tabular}{ll}
    \hline
   Modified Gaia label & SIMBAD codes \\
    \hline
Star & *, Ma*, bC*, sg*, s*r, s*y, s*b, WR*, N*, Psr, Y*O, Or*, TT*, Ae*, out, HH, MS*, Be*, BS*, SX*, gD*, dS*, Ev*, RG*, HS*, HB*, RR*, \\
& WV*, Ce*, cC*, C*, S*, LP*, AB*, Mi*, OH*, pA*, RV*, PN, Pe*, a2*, RC*, SN*, LM*, BD*, Pl, V*, Ir*, Er*, Ro*, Pu*, Em*, PM*, HV*\\
Galaxy & G, LSB, bCG, SBG, H2G, EmG, GiP, GiG, GiC, BiC, IG, PaG, GrG, CGG, ClG, SCG, LeG, PoG \\
Quasar & AGN, SyG, Sy1, Sy2, rG, LIN, QSO, Bla, BLL, LeQ \\
 White dwarf & WD*\\
 Binary star & **, El*, EB*, SB*, BY*, RS*, Sy*, XB*, LXB, HXB, CV*, No*\\
 Unclassified & Cl*, GlC, OpC, As*, St*, MGr, ISM, SFR, HII, Cld, GNe, RNe, MoC, DNe, glb, CGb, HVC,
  cor, bub, SNR, sh, flt, vid, grv, Lev, gLS, gLe,\\
&    LeI, BH, GWE, ev, Rad, mR, cm, mm, smm, HI, rB, Mas, IR, FIR, MIR, NIR, Opt, EmO, blu, UV, X, ULX, gam, gB, mul, err, PoC, ?, reg\\
Ambiguous & Rad, mR, cm, mm, smm, HI, rB, Mas, IR, FIR, MIR, NIR, Opt, EmO, blu, UV, X, ULX, gam, gB\\
 \hline
\end{tabular}
\end{table*}

\clearpage

\section{Classification metric definitions}
\label{appendix-metric}

The classification metrics computed by \textcode{scikit-learn} \citep{scikit-learn} are defined as follows. 
Let $N$ be the number of objects to be classified into $n$ mutually-exclusive classes, 
$(y_{{\rm true},i}, y_{{\rm pred},i})$ be the true and predicted classes, respectively of the $i$-th object,
and 
\begin{equation*}
1(j,k) = \left\{ \begin{array}{cc}
    1 &  j=k\\
    0 & j\neq k
\end{array}
\right.
\end{equation*}
be the indicator function. 
Then
\begin{equation}
\texttt{accuracy}(y_{\rm true}, y_{\rm pred}) = \frac{1}{N} \sum_{i=0}^{N-1} 1(y_{{\rm true},i}, y_{{\rm pred},i}).
\end{equation}
For balanced accuracy, each object is weighted according to the inverse prevalence of its true class, $w_i = 1/(\sum_j 1(y_i,y_j))$ such that:
\begin{equation}
\texttt{bal-accuracy}(y_{\rm true}, y_{\rm pred}, w) = \frac{1}{\sum{w_i}} \sum_i 1(y_{{\rm true},i}, y_{{\rm pred},i}) w_i
\end{equation}

Precision and recall (also known as purity and completeness) are more conveniently defined via set notation, where precision $P(A, B) := \frac{\left| A \cap B \right|}{\left|B\right|} $ and
recall  $R(A, B) := \frac{\left| A \cap B \right|}{\left|A\right|} $.
(For $A = \emptyset $ \textcode{scikit-learn} uses the conventions $R(A, B):=0 $, $P(A, B):=0 $.) 
These quantities can be calculated separately per class and can also be averaged over classes, with the distinction being exactly which sets are used in the calculation.
Let $y_{\rm true}$ be the set of (sample, label) pairs, $y_{\rm pred}$ the set of (sample, label) pairs,
$L$  the set of labels and $S$  the set of samples.
Denote as $y_s $  the subset of  $y$  with sample $s $ and  $y_l$  the subset of  $y $  with label  $l $.
In `macro' averaging, the average is taken over the classes:
\begin{equation}  
\texttt{precision} = \frac{1}{\left|L\right|} \sum_{l \in L} P(y_{{\rm true},l}, y_{{\rm pred},l}) 
\end{equation}
\begin{equation}  
\texttt{recall} = \frac{1}{\left|L\right|} \sum_{l \in L} R(y_{{\rm true},l},y_{{\rm pred},l}) 
\end{equation}
while in `weighted' averaging, each class is weighted by its prevalence in the true sample:
\begin{equation}  
\texttt{precision} = \frac{1}{\sum_{l \in L} \left|y_l\right|} \sum_{l \in L} \left|y_{{\rm true},l}\right| P(y_{{\rm true},l}, y_{{\rm pred},l}) 
\end{equation}
\begin{equation}
\texttt{recall} = \frac{1}{\sum_{l \in L} \left|y_l\right|} \sum_{l \in L} \left|y_{{\rm true},l}\right| R(y_{{\rm true},l},y_{{\rm pred},l}).  
\end{equation}

%%%%%%%%%%%%%%%%%%%%%%%%%%%%%%%%%%%%%%%%%%%%%%%%%%

% Don't change these lines
\bsp	% typesetting comment
\label{lastpage}
\end{document}